\documentclass[useAMS,usenatbib]{mn2e}

\usepackage{graphicx}
\usepackage{natbib}
\usepackage{amssymb}
\usepackage{lscape}

\title[The curious case of NGC~1313]{Evidence of an interaction from resolved stellar populations: The curious case of NGC~1313.}
\author[E. Silva-Villa and S.~S. Larsen]{E. Silva-Villa$^{1}$\thanks{E-mail: esteban.silvavilla@phy.ulaval.ca} and S.~S. Larsen$^{2}$\thanks{E-mail: s.larsen@astro.ru.nl}\\
$^{1}$D\'epartement de physique, de g\'enie physique et d'optique, and \\
Centre de recherche en astrophysique du Qu\'ebec (CRAQ), Universit\'e Laval, Qu\'ebec, Canada\\
$^{2}$Department of Astrophysics, IMAPP, Radboud University Nijmegen, Nijmegen, The Netherlands}
\begin{document}

\date{Accepted 2012 February 22.  Received 2012 February 15; in original form 2011 November 7}

\pagerange{\pageref{firstpage}--\pageref{lastpage}} \pubyear{2011}

\maketitle

\label{firstpage}

\begin{abstract}
The galaxy NGC~1313 has attracted the attention of various studies due to the peculiar morphology observed in optical bands, although it is
classified as a barred, late-type galaxy with no apparent close-by companions. However, the velocity field suggests an interaction with a satellite companion.
Using resolved stellar populations, we study different parts of the galaxy to understand further its morphology. Based on HST/ACS images, we estimated star formation
histories by means of the synthetic CMD method in different areas in the galaxy. Incompleteness limits our analysis to ages younger than $\sim100$Myr.
Stars in the red and blue He burning phases are used to trace the distribution of recent star formation. 
Star formation histories suggest a burst in the southern-west region.
We support the idea that NGC~1313 is experiencing an interaction with a satellite companion, observed as a tidally disrupted satellite
galaxy in the south-west of NGC~1313. However, we do not observe any indication of a perturbation due to the interaction with the satellite galaxy at other locations across 
the galaxy, suggesting that only a modest-sized companion that did not trigger a global starburst was involved.
\end{abstract}


\section{Introduction}
\begin{figure*}
\centering
\includegraphics[width=150mm]{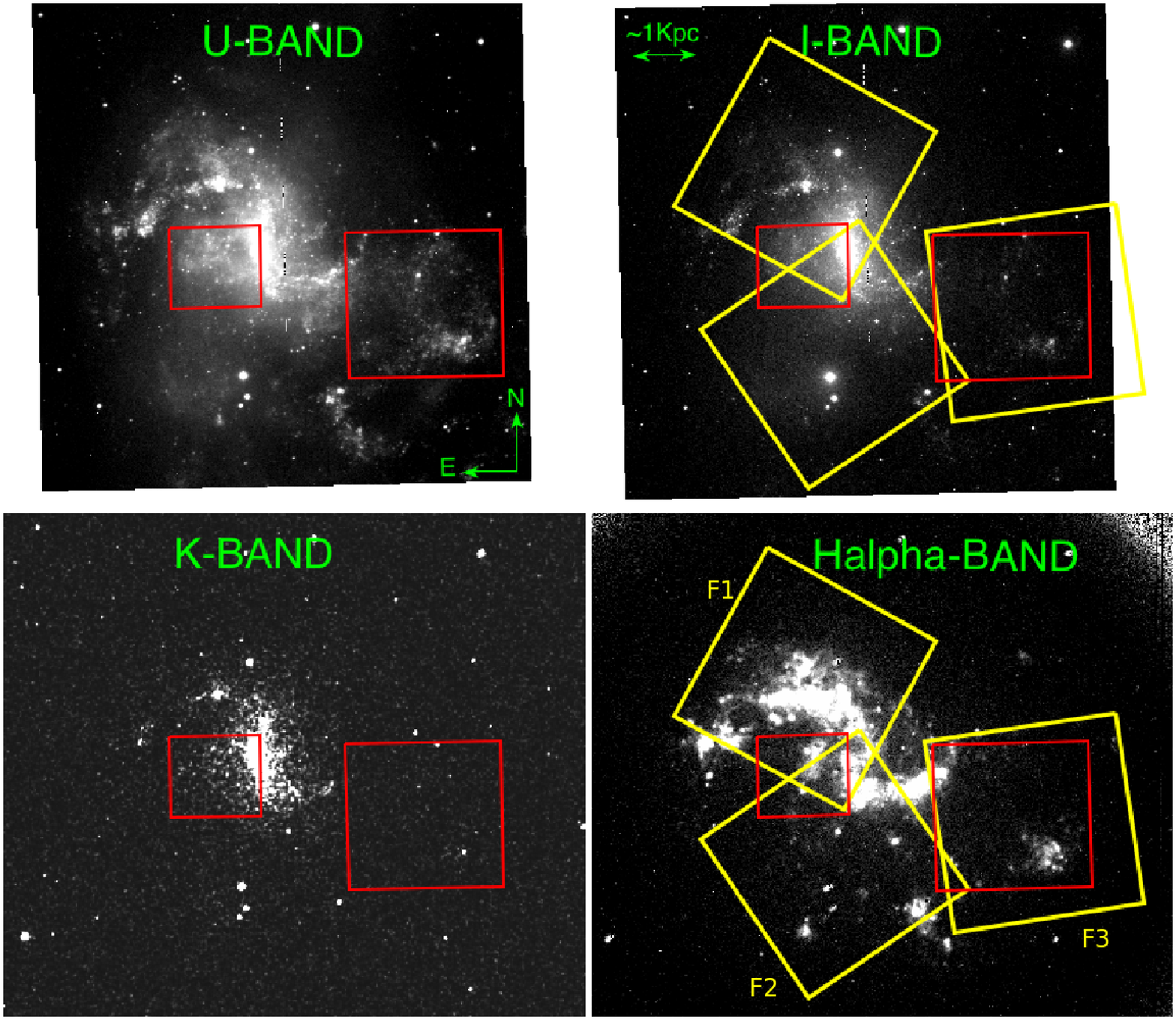}
\caption{NGC~1313 observed through different filters. Images have arbitrary intensity scales, however, they all have matched WCS coordinates. Original sources of the images are: {\em U and I}: Las Campanas Observatory \citep{kuchinski00}; {\em K}: 2MASS Survey \citep{jarrett03}; and {\em H$\alpha$}: CTIO \citep{hoopes01}. The images were taken from NED data base. Red boxes illustrate the areas discussed in this paper. The yellow boxes illustrate the three ACS pointings used during this study and are labeled as F1, F2 and F3.}
\label{fig:difbands}
\end{figure*}

The late type barred spiral galaxy NGC~1313 has been classified as a galaxy in transition between SBm and SBc by \citet{devaucoulers63}.
After observations in different bands, the peculiar morphology (see Fig. \ref{fig:difbands}), the indications
of a possible interaction with a satellite galaxy \citep[e.g.][]{sandage79} and the particular shape of the arms, called the attention of many studies. 

Figure \ref{fig:difbands} shows NGC~1313 in the bands U, I, K, and H$\alpha$. Looking at the bar of the galaxy (see left small red boxes in Fig. \ref{fig:difbands}) some issues call the attention: 
(1.) the bands U and H$\alpha$ can trace recent star formation, and show to be enhanced in the bar and close to its east side;
(2.) the bands I and K can trace intermediate/old populations (although, also relatively young) which appear to be stronger in the bar of the galaxy compared to the arms;
(3.) the H$\alpha$ image shows a region next to the bar (east side) that does not appear to be connected with the bar nor with the arms;
(4.) a detached region to the south-west is particularly prominent in the U-band image; and  
(5.) the bar and spiral arms are embedded in a faint, elongated region of diffuse light running from the south-east towards the north-west, visible in the U-band image 
\citep[see also][]{ryder95}. The surface brightness of this emission appears somewhat higher towards the southern part of the images in Fig. \ref{fig:difbands}.

Due to the observed unusual morphology \citep[e.g.][]{marcelin79,blackman81}, studies of this galaxy claimed for a displacement
of the dynamical center of rotation with respect to the optical center. However, accurate
measurements of the velocity field was presented by \citet{peters94} using HI and 1.4 GHz observations,
showing that the galaxy presents a smooth velocity field in the regions close to the bar, contrary to the results presented by previous
studies, concluding that there is no such displacement. 

Comparing models and observations of velocity fields, \citet{marcelin82} estimated values of the M/L ratio.
Marcelin \& Athanassoula suggested lower values of M/L in the bar (3-10 times smaller) compared to the inner disc.
This result, combined with their observation of a displacement in the bar (i.e. of the dynamical center of rotation with respect to the optical center) was suggested as two different populations,
one in the bar of the galaxy and a second one in the inner disc. \citet{walsh97} suggest a possible enrichment in the bar of the galaxy
based on (slightly higher, $\sim0.2$ dex) measurements of O/H, compared to the disc.
\citet{molla99} used a multiphase chemical evolution model trying to reproduce
the flat abundance gradient observed in NGC~1313 \citep{walsh97}. Their model was not able to recover
at the same time the abundance distribution, gas density and the star formation profiles obtained by observations, which 
makes the galaxy a particular and special case to study.

From the U- and H$\alpha$ images in Fig. \ref{fig:difbands}, there is an indication of recent star formation activity in the active region in the south-western part of the galaxy.
\citet{sandage79} suggested that this is the remnants of a tidally disrupted companion galaxy. The possibility
of a satellite companion in process of interaction with NGC~1313 was also suggested by \citet{blackman81}.
Blackman found that the velocity field of NGC~1313 was inconsistent with pure rotation profiles, which is interpreted by Blackman
as an indication of the presence of a satellite companion. Velocity fields obtain by \citet{peters94}
show a clear disturbance over the south-west area of NGC~1313, suggesting an interaction between two galaxies. 
Observing the same region through the bands I and K it suggests the presence of previous populations in the region, 
which do not seem to be connected to the southern arm of NGC~1313, possibly associated with the satellite galaxy. 
However, I and K bands can also trace red super giants (i.e. relatively young stars). 

\citet{larsen07} studied resolved stellar population and (massive) star clusters in NGC~1313 using the same observations used in this paper.
Larsen et al. suggested an increase in the star formation history over the south-west region of the galaxy (region covered by the right big red 
boxes in Fig. \ref{fig:difbands}). The possible cause of the increase
observed by Larsen et al. could be related with the tidal interaction observed by \citet{marcelin79,sandage79,blackman81} and
\citet{peters94}.

On a global scale, both observations and theory indicate that 
 interactions and/or galaxy mergers influence the evolution and 
 morphology of the galaxies involved, and may trigger enhanced
 star formation \citep[e.g.][]{barnes99,schweizer00}. Therefore, if 
 NGC 1313 underwent an interaction as suggested by previous studies, 
 we may expect that there should be an increase in the star
 formation rate. Observations of the stellar populations and star
 formation histories in different regions of NGC 1313 may thus
 help answer the question whether an interaction took place in the
 recent past and triggered (local or global) bursts of star formation.
 
Based on Hubble Space Telescope observations, we study NGC~1313, looking to understand further its morphology.
Thanks to the superb resolution of the HST, we can use the resolved stellar population as a fingerprint of past events.
We have observations of three main regions over the galaxy,
covering the northern and southern arms, the bar of the galaxy, and the region that may have been disturbed by an interaction with a satellite companion (see yellow boxes in Fig. \ref{fig:difbands}).
The location of our observations will allow us to test and understand further the morphological features of NGC~1313.
If there were any interaction and/or merger in the past, the resolved stellar populations may hold further clues to the nature of these events.
We present our observations and photometry in Sect. 2. Using the colour-magnitude diagrams
for the resolved stars, we devoted Sect. 3 to see what information can be extracted from there. We study the star formation
histories of the different fields observed and for different group of stars in Sect. 4. We discuss our results and give a summary and conclusions in
Sects. 5 and 6.

\section{Observations and photometry}
NGC~1313 was observed using the {\em Advanced Camera for Surveys} (ACS)
onboard the the {\em Hubble Space Telescope} (HST) as part of Cycle 12 in 2004.
The instrument has a resolution of $0\farcs05$ per pixel. With a distance modulus of  $28.08$ \citep[$\sim 4$Mpc,][]{mendez02}, 
1 pixel corresponds to $\sim$1 pc.
Optical bands were covered using the filters F555W ($\sim$V), and
F814W ($\sim$I), with exposure times of 680 sec for the V-band, and 676 sec for the I-band.
Three main regions were targeted covering the north-east, center-east, and south-west (see Fig. \ref{fig:difbands}).

Standard STScI pipeline was used for the initial data processing. ACS images were drizzled using the multidrizzle 
task \citep{koekemoer02} in the STSDAS package in IRAF. We use the default parameters, but disabling the
automatic sky subtraction. A full description of our procedures for object detection over the images can be found in \citet{silvavilla10}.


\subsection{Field stars photometry}
Stellar photometry was done by means of point-spread function (PSF) fitting photometry due to crowding.
Visually inspecting the FWHM (with {\it imexam} in IRAF)
a set of isolated bona-fide stars were used to construct the PSF
in each band. PSF photometry was done with {\it DAOPHOT} in IRAF.

Our PSF-fitting magnitudes are corrected to a nominal aperture radius of $0\farcs5$, following 
standard procedures. From this nominal value to infinity, we apply the corrections in \citet{sirianni05}.
HST zero-points\footnote{www.stsci.edu/hst/acs/analysis/zeropoints/\#tablestart} have been applied to the PSF magnitudes 
after applying aperture corrections. The zero-points used in this work are 
$ZP_V=25.72$ and $ZP_I=25.52$ magnitudes. Typical errors of our photometry do not change 
dramatically from the ones in \citet[][see its Fig. 2]{silvavilla10}, and we refer the reader to that paper for more details
about our photometry procedures. 

\begin{figure*}
\includegraphics[height=130mm,width=180mm]{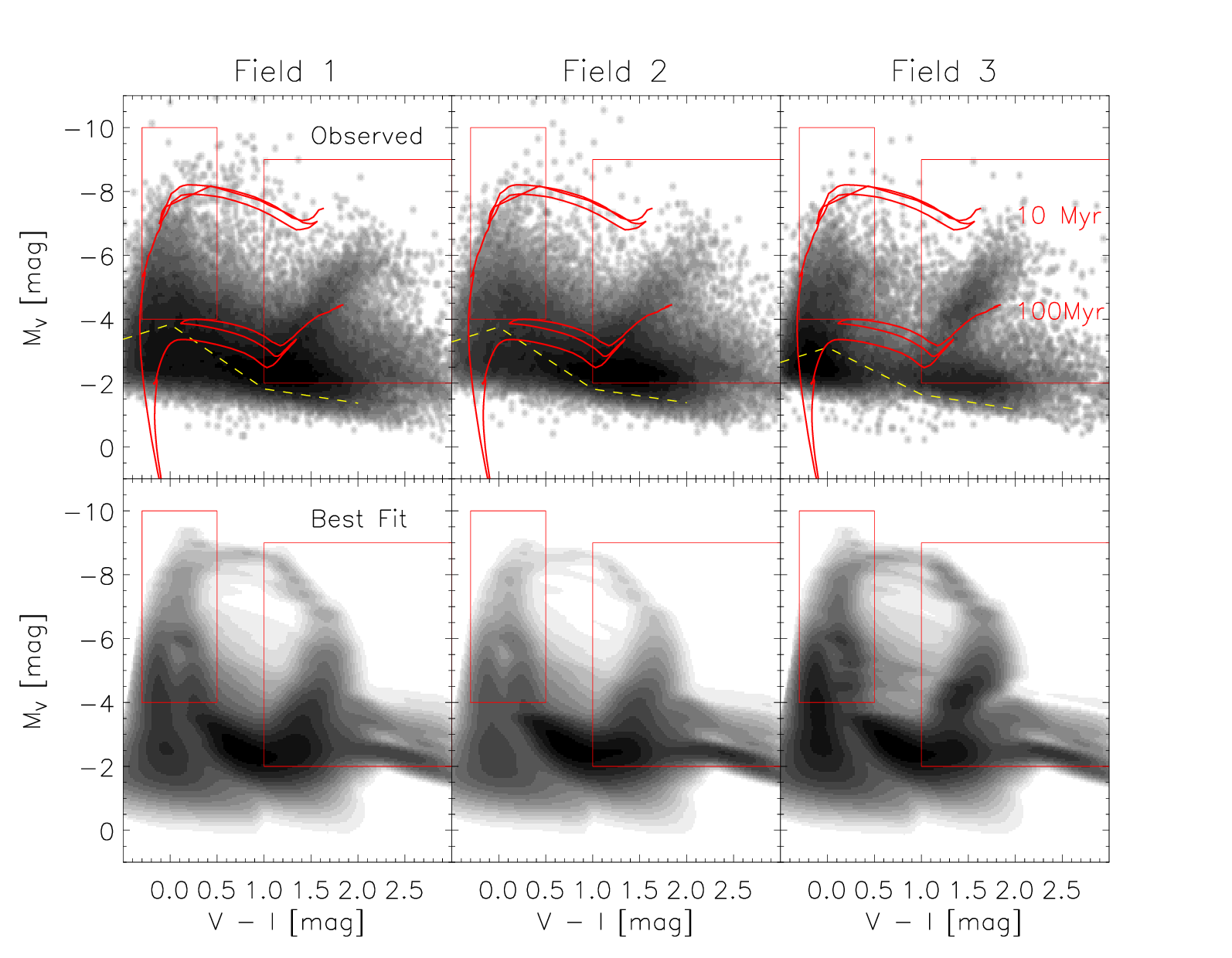}
\includegraphics[width=180mm]{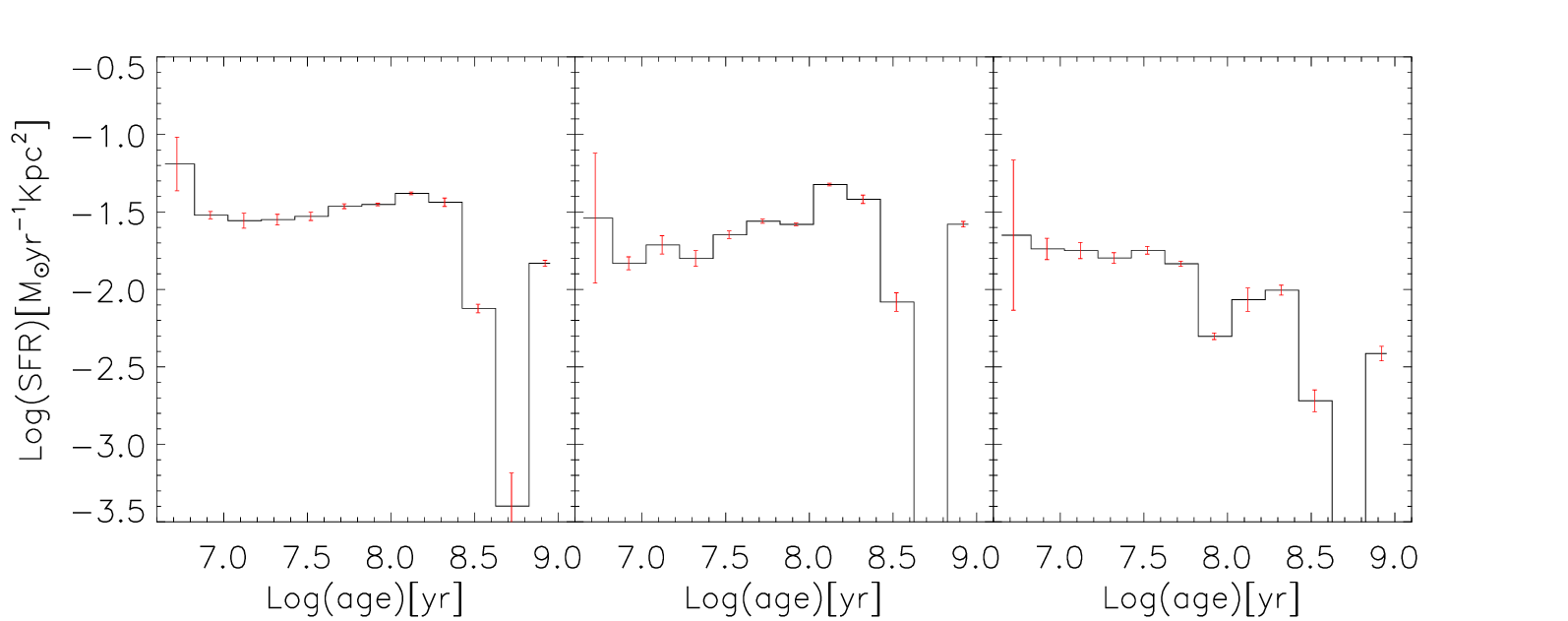}
\vspace{0.2in}
\caption{{\em First row}: CMDs for the three fields observed over NGC~1313. Dashed lines represent the colour-dependent 50\% completeness limits.
	     Red lines represent isochrones at 10 and 100 Myr. Red rectangles are the areas used for the SFH fitting. 
	     {\em Second row}: Best fit of the SFH. {\em Third row}: SFHs estimated in this paper. 
	     Error bars are Poissonian, estimated following the bootstrapping method done in \citet{silvavilla10}.}	
\label{fig:cmdsfh}
\end{figure*}

The final V-I colour-magnitude diagrams (CMD) for the stars in the three fields are presented in Fig. \ref{fig:cmdsfh} (first row). 
The colour-dependent 50\% completeness limits for NGC~1313 were estimated in \citet{silvavilla11}, and are overplotted
in Fig. \ref{fig:cmdsfh} as dashed lines.

\begin{figure*}
\centering
\includegraphics[scale=0.7]{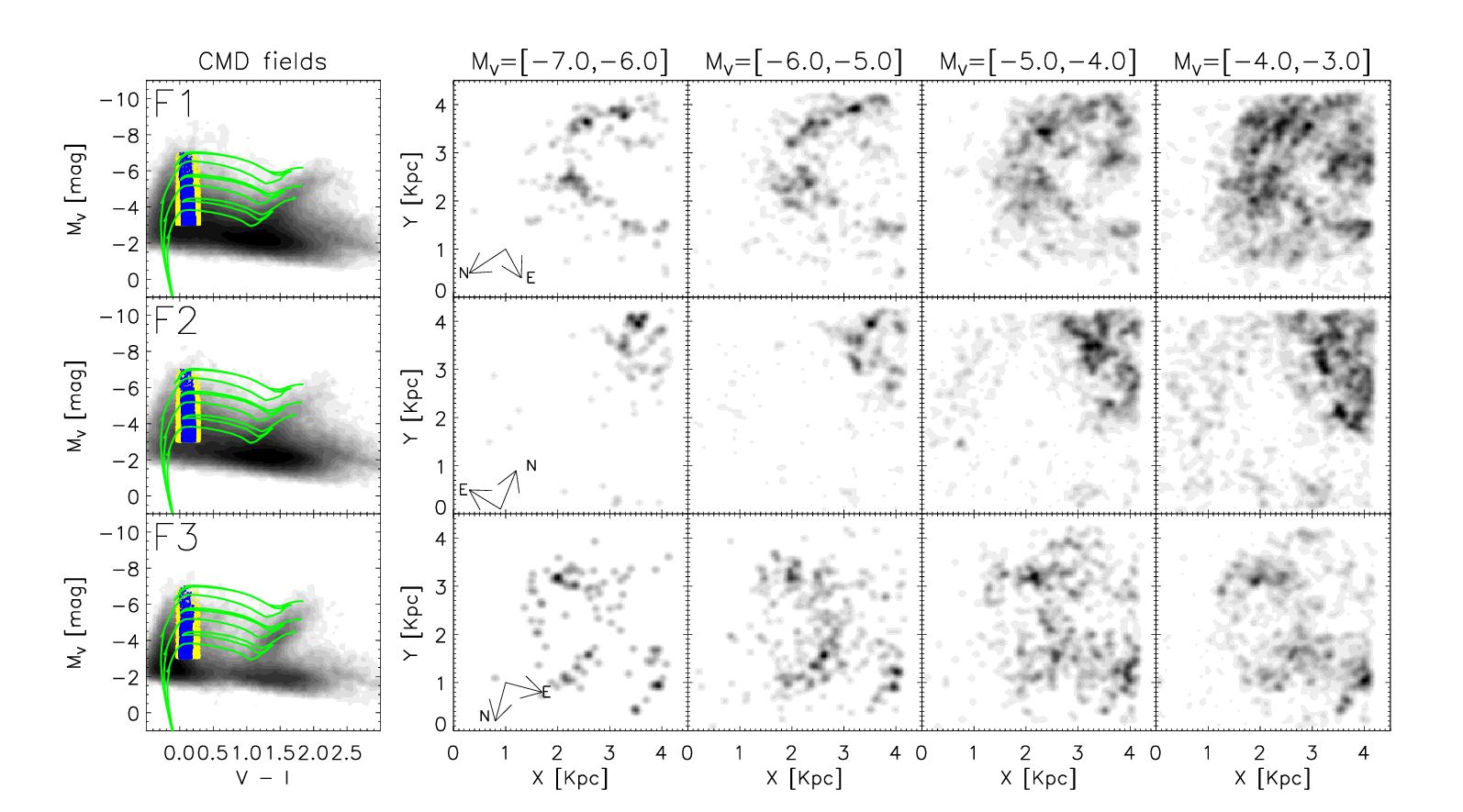}
\includegraphics[scale=0.7]{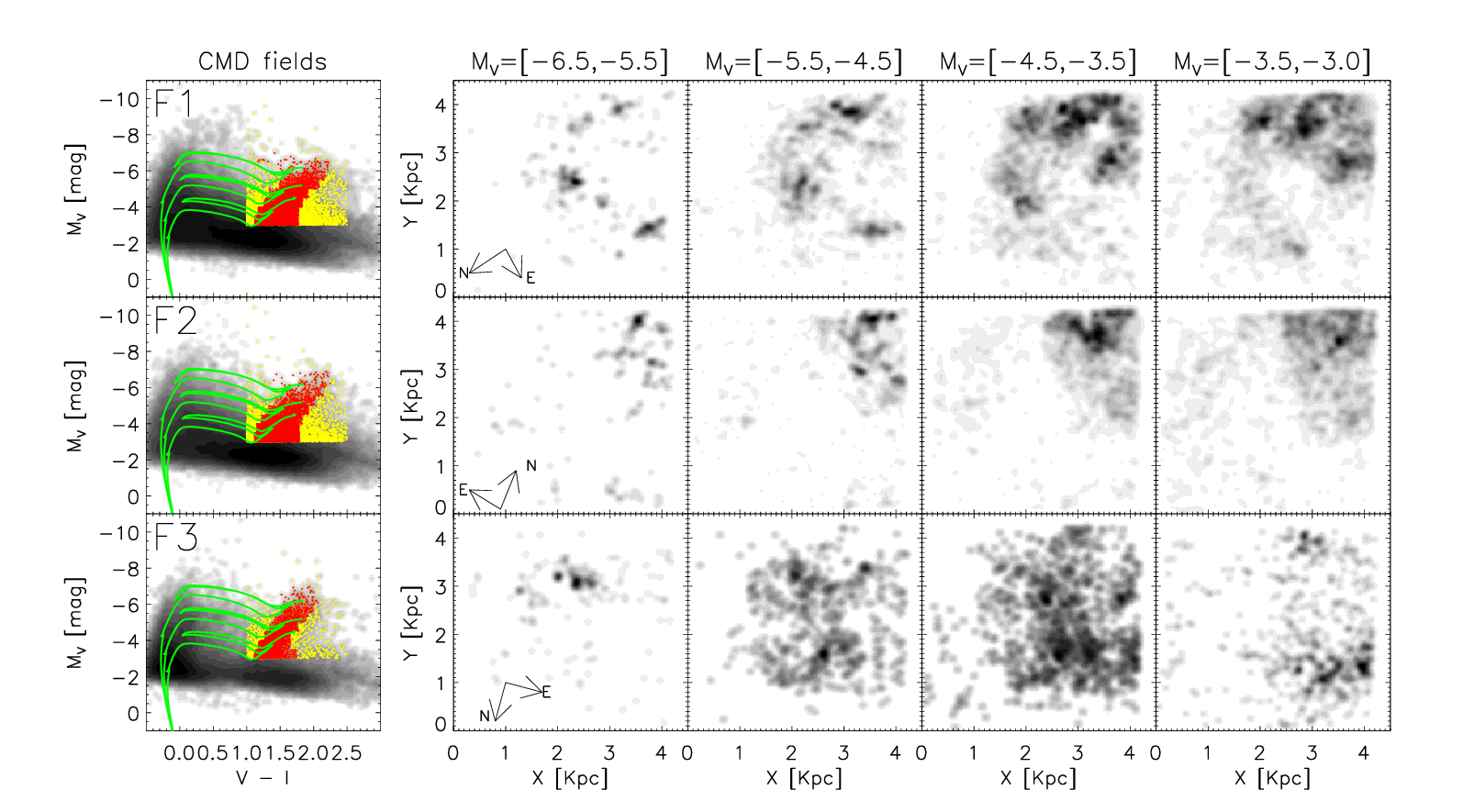}
\caption{Blue and red He burning phases and how are their spatial distributions using different magnitude bins which are correlated with the isochrones plotted. 
	     Padova 2008 \citep{marigo08} isochrones for LMC-like metallicity \citep{walsh97} are overplotted with at $log(\tau)$[yr]$=[7.3,7.6,7.9]$.}
\label{fig:heblue}
\end{figure*}

\section{What colour-Magnitude diagrams can tell us}
A direct comparison among CMDs (see first row in Fig. \ref{fig:cmdsfh}) starts to indicate some differences across the galaxy.
Observing the CMD of field 3, we observe a decrease in the number of stars between the phases red super giants (RSG) and red giant branch (RGB), which
is not observed in the CMDs of the other two fields. This ``gap" (observed at $V-I\approx1.3$ and $-3.5 \le$ M$_V \le -2.5$) suggests that the star formation history of field 3 is different
in comparison with the other two fields. Also, the density of stars in the young stages (i.e. main sequence) is different among
the fields, being relatively higher in field 3 compared to field 2. We try to observe further differences
through observations of He burning phases among the fields and their spatial distribution.
%

Stars in the red and blue He burning phases (hereafter referred as R$_{He}$ and B$_{He}$ phases) are well age-dated by theoretical isochrones, and a comparison
between these two group of stars and isochrones at different ages can (roughly) indicate how the recent star formation is distributed over the galaxy.
Stars in the He burning phases have been used to trace star formation in other galaxies \citep[e.g.][for Sextant A]{dohmpalmer97}, however, other features in
the CMDs such as the main sequences turn-off and/or sub-giant branches, are being used for the same purpose \citep[see the review by][and references therein]{gallart05}.
Due to the distance of NGC~1313, we can not use all the possible features in the CMD to estimate/trace star formation histories, many of them fall below our completeness limits, so we
focus the study in this section on the use of the R$_{He}$ and B$_{He}$ phases.

We selected stars that could be in their He burning phases between the magnitudes intervals $-3\ge$ M$_V\ge-7$ and the colour ranges $1\le$ (V-I) $\le2.5$ and $-0.05\le$ (V-I) $\le0.3$
for the red and blue sides, respectively. These two colour ranges were established by visual inspection, i.e. the B$_{He}$ phases 
must be parallel to the main sequence stars, and on their right, while for the R$_{He}$ phases the regions are largely spread over the parameter space, so we used a larger colour range.
The final sample of stars in the R$_{He}$ and B$_{He}$ side was obtained estimating the mean value of the color distribution and the
1$\sigma$ deviation every 0.5 magnitudes in $M_V$. We illustrate our results in Fig. \ref{fig:heblue}, where stars in yellow are stars outside the
1$\sigma$, while red and blue represent the two He burning phases respectively. Overplotted are three isochrones representing ages at 
$\tau=[10^{7.3},10^{7.6},10^{7.9}]$yr ($1^{st}$ column in Fig. \ref{fig:heblue}), assuming LMC-like metallicity 
\citep{walsh97} and using Padova 2008 isochrones \citep{marigo08}.
To see how stars at different ages are distributed over the galaxy, we used stars close to each
isochrones, separating them in the magnitude bins M$_V=[-6.5,-5.5,-4.5,-3.5,-3]$ and M$_V=[-7,-6,-5,-4,-3]$ for the R$_{He}$ and B$_{He}$,
respectively (see columns $2^{nd}$ to $5^{th}$ in Fig. \ref{fig:heblue}). These bins roughly correlate with the three isochrones overplotted.
Based on the spatial distributions of R$_{He}$ and B$_{He}$ at different magnitudes (i.e. different ages), we observe that:

\begin{itemize}

\item {\em Field 1}: The first two panels (from left to right for R$_{He}$ and B$_{He}$) show that the brighter stars are concentrated in the northern arm of NGC~1313, 
			      however, the density of B$_{He}$ appear to be larger as approaching the fainter magnitudes (last two panels on the right).
			      This larger amount of stars in the fainter magnitudes for the B$_{He}$ could be a consequence of contamination from main sequence
			      stars, which are harder to separate at fainter magnitudes. The region indicated in Fig. \ref{fig:difbands} with an 
			      enhancement in H$\alpha$ (small red boxes in the figure),
			      is located around $(X,Y)\approx(4,3)$Kpc. We observed how this region starts to be more visible when moving to the fainter magnitudes in the R$_{He}$
			      panels, indicating a possible increase in the star formation in the past. Also observed in the R$_{He}$ panels is an increase in the number of stars at the
			      end of the arm ($(X,Y)\approx(4,1.5)$Kpc, brighter panel). However, in a general view, there is an apparent increase.

\item {\em Field 2}: The spatial distributions of the stars appear to be separated into two regions, from $\sim2$Kpc 
			       to the left and to the right, assuming a vertical division from the center of the field of view. 
			       It is important to note that there is no apparent formation of stars at the center of this field. These two statements are more
			       clear in the B$_{He}$ than in the R$_{He}$ panels. Nevertheless, the concentrations are higher in the right side, where the bar is located.
			       The region on the left resembles a stream of stars, however, this region is not connected with the arms of the galaxy nor 
			       with the bar.

\item {\em Field 3}: In the youngest bin we see a region of active/very recent star formation towards the southern part of the field (at $(X,Y) \approx (2.5,3.5)$Kpc). 
			        In the second bin we see a fairly smooth distribution of stars, many of which are found within an elongated structure with a radius of $\sim 1$ 
			        kpc that is roughly centered on the ACS field of view. The third bin shows an overdensity of stars towards the northern part of this structure 
			        (at $(X,Y) \approx (3,1.5)$Kpc) and in the oldest bin most of the stars are associated with this overdensity. It thus appears that there has been a 
			        progression of star formation from the northern part of the field towards the south over the past ~80-100 Myr.

\end{itemize}

\begin{landscape}
\begin{figure}
\centering
\includegraphics[scale=0.85]{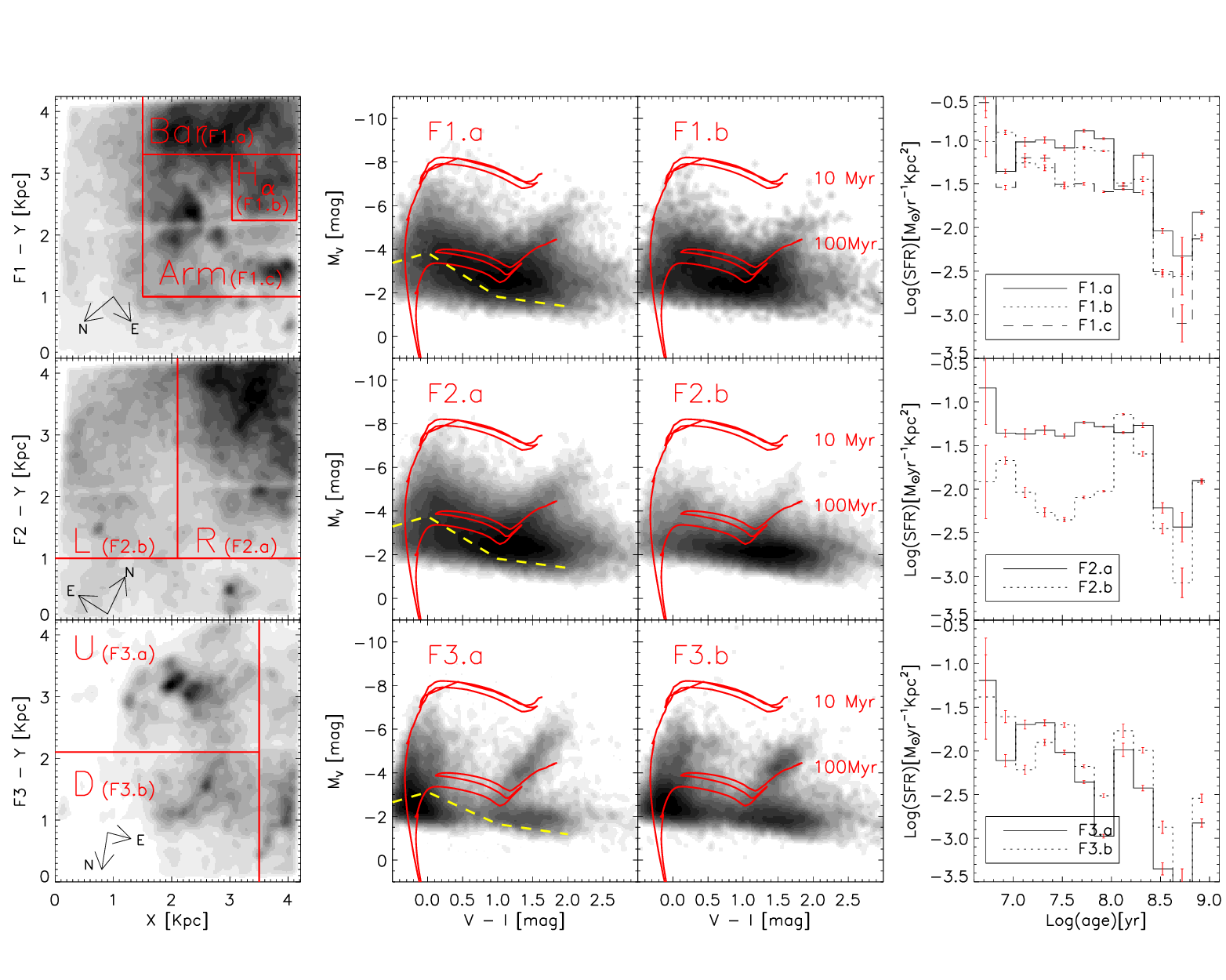}
\caption{{\em Columns 1}: Density of stars over the three fields observed. The red lines indicate the subdivisions used to study different areas (see text for more details). 
	       {\em Column 2 and 3}: CMDs of the different areas. Red lines represent isochrones at 10 and 100 Myr \citep{marigo08}, 
	       and yellow dashed line is the 50\% color-dependent completeness
	       estimated by \citet{silvavilla11}. {\em Column 4}: Star formation histories. Error bars are Poissonian and were estimated following the bootstrapping method described in
	        \citet{silvavilla10}.}
\label{fig:sfhall}
\end{figure}
\end{landscape}

			      
Based on the distribution of the density of stars observed in the fields, we sub-divided our fields to observe
changes in the SFH that can give more information on the evolution of this galaxy:
\begin{itemize}
\item Our field 1 covers three main regions: \\
(1.) {\em the bar}, located in our images between $(X_i,Y_i)=(1.5,3.3)$ and $(X_f,Y_f)=(4.2,4.2)$ Kpc;
(2.) {\em the ''H$\alpha$ region''}, located in our images between $(X_i,Y_i)=(3.0,2.2)$ and $(X_f,Y_f)=(4.1,3.3)$ Kpc; and
(3.) {\em the northern arm}, located in our images between $(X_i,Y_i)=(1.5,1.0)$ and $(X_f,Y_f)=(4.2,1.5)$ Kpc, removing
the region covered by the {\em H$\alpha$ region}.

For the rest of the paper, we will refer to these three regions as: F1.a, F1.b and F1.c, respectively.

\item The field 2 covers two main regions: \\
(1.) {\em the region R}, located in our images between $(X_i,Y_i)=(2.1,1.0)$ and $(X_f,Y_f)=(4.2,4.2)$ Kpc; and
(2.) {\em the region L}, located in our images between $(X_i,Y_i)=(0.0,1.0)$ and $(X_f,Y_f)=(2.2,4.2)$ Kpc.

For the rest of the paper, we will refer to these three regions as: F2.a and F2.b, respectively.

\item The last field, field 3, covers two main regions: \\
(1.) {\em the region U}, located in our images between $(X_i,Y_i)=(0.0,2.1)$ and $(X_f,Y_f)=(3.7,4.2)$ Kpc; and
(2.) {\em the region D}, located in our images between $(X_i,Y_i)=(0.0,0.0)$ and $(X_f,Y_f)=(3.7,2.1)$ Kpc.
Figure \ref{fig:sfhall} presents the density of stars of each field and the subdivision used for further analysis.

For the rest of the paper, we will refer to these three regions as: F3.a and F3.b, respectively.

\end{itemize}

CMDs for each of the subsections were created to see if they show any differences (see columns 2 and 3
in Fig. \ref{fig:sfhall}). Color-dependent completeness limits estimated by \citet{silvavilla11} and isochrones at 10 and 100 Myr \citep{marigo08} are overplotted. 
We do not include the CMD of the the region F1.c (northern arm) because it does not
present any significant differences in comparison to the other regions in the same field. Between the regions F1.a and
F1.b located next to it (identified in Fig. \ref{fig:difbands}, small red boxes), we do not see
major differences. Both regions present a spread in absolute magnitude of the red and blue core He-burning stars, implying an age spread.
However, the R$_{He}$ phases of F1.a reach magnitudes brighter than
the F1.b area (by $\sim$1 magnitude), suggesting differences in the star formation activity. Field 2 presents two regions, F2.a and F2.b. The CMDs of these regions
present some differences. First, the difference in the density of stars in both areas is clear, where F2.a has a much larger number
of stars. This is expected, because part of the bar of the galaxy is covered by this region, as well as part of the  southern arm. 
Second, the tip of the main sequence and the R$_{He}$ phases
is much brighter in the F2.a than in F2.b, indicating that F2.a region has a higher star formation rate at recent times in comparison with the F2.b region.
Also, the high density of stars in the RGB/AGB phases observed in the F2.b indicate a very old population, and the low density of stars in the main sequence
indicates that either there is no star formation occurring at the moment or that the star formation rate is very low. The CMDs
in the field 3, regions F3.a and F3.b, also show interesting characteristics. In the F3.a region, the separation between the
main sequence and the B$_{He}$ phases is clearly visible, something not seen in any of the other CMDs presented.
It also presents a very narrow sequence of R$_{He}$ stars, which reaches
magnitudes of M$_V\approx-7$. The region F3.b presents also a large population of stars in the main sequence, however
the population of stars in the RGB/AGB has a higher number of stars compared with the F3.a. The location of F3.b is close to the southern 
arm of NGC~1313, so the possibility of a contamination of stars that do belong to the arm of the galaxy and stars that do not follow the
arm distribution can cause this effect. Also important to note is that the peak of the
R$_{He}$ phases in the F3.b region is $\sim0.5$ magnitudes fainter in comparison with the F3.a region, and 
the range in M$_V$ magnitude of the R$_{He}$ stars is narrower in region F3.b, indicating a smaller age range.


\section{Star formation histories}
We employ the synthetic CMD method \citep{tosi91} to obtain the star formation histories (SFH).
Our IDL-based program for the estimation of SFH was created and tested on the galaxy NGC~4395 \citep{silvavilla10}. In a further
study, \citet{silvavilla11} studied the SFH of NGC~1313 (as a whole) as part of a sample of five galaxies and found a
$\Sigma_{SFR}$=11.26$\times$10$^{-3}$ M$_{\odot}$yr$^{-1}$Kpc$^{-2}$. In comparison with the rest of the sample in
Silva-Villa \& Larsen, NGC~1313 is the second most active galaxy.

The parameters used to estimate the SFH in this paper are: a distance modulus of 28.08, Magellanic Clouds-like metallicities 
\citep[$Z=(0.004,0.008)$\footnote{There is not constrain on the metallicities for this galaxy, and as shown in \citet{silvavilla11}, 
the use of more than one metallicity drives to better fits.},][]{walsh97}, a binary fraction of
0.5 with a mass ratio between 0.1-0.9 (assuming a flat distribution and no binary evolution), a foreground extinction of A$_b$=0.47 \citep{schlegel98}, 
and the colour combination V-I. We adopted Padova 2008 isochrones \citep{marigo08}. The results were normalized by the areas covered, 
having then the surface star formation rate density ($\Sigma_{SFR}$[M$_{\odot}$yr$^{-1}$Kpc$^{-2}$]).

The best fits obtained and the estimated SFHs for individual fields are presented in Fig. \ref{fig:cmdsfh} (second and third row). In general, the SFH for the three fields
are very similar, showing a constant star formation over time. However, there is an apparent increasing trend observed in the field 3 
and there is a clear dip in the SFH at $\sim10^8$ years, corresponding to the gap at the foot of the R$_{He}$ sequence already pointed out above.
Nevertheless, our completeness limits are very close to the isochrones at $10^8$ yr, so we do not assign any physical meaning to the ``well" observed at 
$\sim10^{8.7}$ yr, which is visible in the three SFHs.

Due to the differences observed in the CMDs in the previous section, we estimate SFHs for each of the subsections over our fields.
Estimations of the mean SFRs, and the respective standard deviation of the mean, for the full three fields and the subareas in each field are
presented in table \ref{tab:sfhs}.

{\em The regions F1.a, F1.b and F1.c}: SFH estimations of the three regions covered in our field 1 do not show a significant
difference for ages between $\sim$10 to $\sim$40 Myr, however, we observe a higher value ($\approx$0.4 dex) from $\sim$40 up to 100 Myr between the 
regions F1.b and F1.c (see column 4 in Fig. \ref{fig:sfhall}), even F1.a shows a higher value at this age range compared to F1.c.
The F1.b region in the field appears to be disconnected from the bar and the arms of the galaxy, nevertheless the SFH observed in Fig. \ref{fig:sfhall}
shows no significant difference. It is important to note that the F1.b region presents a higher SFR between 
$\sim$40 up to 100 My, showing even a larger mean value (see table \ref{tab:sfhs}) in comparison to the other two regions observed in that field. 

{\em The regions F2.a and F2.b}: The comparison between these regions shows a more intriguing result. As observed in the density plots of the 
stars and the CMDs for these two regions in the field 2 (see column 1, 2, and 3 in Fig. \ref{fig:sfhall}), the star formation history in the 
bar is different than in the apparent stream of stars located in the F2.b region. The peak in the SFH observed
at $\sim10^{8.2}$ yr is very close to our completeness limits, so we do not assign any physical meaning. 
Nevertheless, there is a difference of about $\sim$1 dex between the star formation rates for the two regions at younger ages ($\le$100 Myr).

{\em The F3.a and F3.b regions}: The SFH for these regions shows particular features. First, there is a clear increase in the SFR for both regions
between $10^{7.4}$ to $10^{7.8}$ yr. Second, between $10^{7.1}$ and $10^{7.4}$ yr there is an increased separation in the SFHs by a factor of $\sim$0.5 dex, being more pronounced
at the young side. And third, there is a dip at $10^{7.9}$ yr, which was suggested already form the CMDs (see Fig. \ref{fig:sfhall}), where there
is an apparent gap between the RSG and RGB. An increase in the SFH for NGC~1313 in the same region has been observed previously by \citet{larsen07}, using the same dataset
used in this work. Larsen et al. suggested the increase of the star formation based on two results: (1.) there is an increase in their
estimations of the star formation rate, and (2.) they estimate the fraction between the number of "old"  over "young" massive clusters (i.e. M$\ge 5000$ M$_{\odot}$),
showing to have a higher value for the field 3 compare to the other two areas observed. We concluded that the increase in the SFH for the third field is a physical effect acting over
the last $\sim$100 Myr.


\begin{table}
\centering
\caption{Mean values of $\Sigma_{SFR}$ [$\times 10^{-3}$M$_{\odot}$yr$^{-1}$Kpc$^{-2}$] for the three fields and sub fields. Estimations are
	      between 10 and 100 Myr.}

\begin{tabular}{c c c}
\hline \hline
Field/Subfield & Mean & STDDEV  \\  \hline
\multicolumn{3}{c}{\underline{Full field}} \\
F1 & 30.9 & 3.2 \\
F2 & 21.0 & 5.3 \\
F3 & 14.9 & 5.1 \\ \hline
\multicolumn{3}{c}{\underline{Field 1}} \\
F1.a & 54.8 & 12.1 \\
F1.b & 60.9 & 28.7 \\
F1.c & 42.0 & 16.0 \\ \hline
\multicolumn{3}{c}{\underline{Field 2}} \\
F2.a & 51.7 & 16.8 \\
F2.b & 9.2 & 4.1 \\ \hline
\multicolumn{3}{c}{\underline{Field 3}} \\
F3.a & 12.4 & 7.9 \\
F3.b & 11.9 & 5.3 \\ \hline

\label{tab:sfhs}
\end{tabular}
\end{table}




\section{Discussion}
The patchy shape (observed through different wavelengths) in the northern arm calls the attention for being an unusual feature for an isolated spiral galaxy.
\citet{peters94} found a loop of hydrogen around NGC~1313, which intersects the location of the satellite companion in the south-west of NGC~1313. The result was interpreted by Peters et al. as
the possible orbit followed by the (disrupted) companion galaxy. The orbit suggested by the authors passes from the northern arm and goes back down to the southern arm,
showing to have an almost circular orbit with a circular velocity $\sim105$ Km/s. Assuming a circular orbit with radii 8 Kpc, the time to cover half orbit
around NGC~1313 is $\sim237$ Myr. We do not observe any physical features in the CMDs nor in the SFH of the field 1 that can suggest the passage of an object close to
the northern arm. However, our measurements are constrained to the last 100 Myr due to incompleteness, so if the satellite galaxy
passed by the northern arm, our data might not be enough to see the effects induced by this process. Nevertheless, the SFH
observed in field 3 in Fig. \ref{fig:sfhall} ramps up around 100 Myr ago, which could be the results of the close encounter with NGC~1313 with the satellite.

To trace regions of recent star formation, H$\alpha$ imaging is widely used. Observing Fig. \ref{fig:difbands}, the area
close to the bar (see small red boxes in the figure) suggested recent activity,
which, visually, do not appear to be connected with any of the arms of the galaxy, nor
with the bar. We estimated the SFH of that particular region, and found a slight difference among the bar, the northern
arm and this region ($\le$0.4 dex the largest difference, observed at $\sim$10$^{7.8}$ Myr), indicating that the 
whole area has been forming stars at (more or less) constant rate (see values in table \ref{tab:sfhs}
for F1.a, F1.b, and F1.c, and column 4 in Fig. \ref{fig:sfhall}). These particular features suggest that the formation 
history of the region F1.b is similar when compared with the 
surrounding environment, however leaves a question open regarding its dynamical process(es). 

Optical and H$\alpha$ images indicate star formation activity in the bar of the galaxy. The 
distribution of stars in the R$_{He}$ and B$_{He}$ phases (see Fig. \ref{fig:heblue}), and the density maps (see Fig. \ref{fig:sfhall}), 
covered by our observations close to the bar (Field 2) suggests two different group of stars.
This has been confirmed based on separated SFH estimations for the regions F2.a and F2.b, which
present a difference of $\sim1$ dex for ages $\le100$ Myr ago (see Fig. \ref{fig:sfhall}). The result observed
in the SFHs was expected based on the difference in the amount of stars and their spatial distribution,
which show a large difference between these two regions. It remains a puzzle the particular spatial distribution of the stream of stars
in the region F2.b, because those stars do not appear linked with either of the two arms of the galaxy, not with the bar either.

The estimations by \citet{marcelin82} of the M/L ratio in the bar compared to the estimations in the inner part of the disc showed to be 3-10 times smaller,
which was interpreted by the authors as an indication of two different populations. A similar result was presented
by \citet{walsh97}, where they observe (slightly, only $\sim0.2$ dex) higher levels of O/H in the bar compared to the arms, suggesting an enrichment process.
Our results suggest two different populations, one being the bar of the galaxy, while there is a second one, observed as a
stream of stars, which have a clearly different SFH (see the discussion in the previous section on the regions F2.a and F2.b). 


\subsection{Tidally disrupted satellite companion}
The work done by \citet{blackman81} and \citet{peters94} on the velocity field of NGC~1313 suggested an interaction
with a satellite companion at the south-west region of the galaxy. \citet{larsen07}
observed an increase in the SFH at the same location. Our observations of this region
have shown that the star formation (as a whole) presents lower levels compared to the ones obtain for the other two regions covered, however
with an indication of an increase over time (see F1, F2, and F3 in table \ref{tab:sfhs}). Separating our field 3 into two different regions (F3.a and F3.b) we have shown that
both are passing through an increase in their star formation history over the last 100 Myr, up to present times. 
This increase in the SFR may have been induced by tidal forces due to an interaction.
We also note from Fig. \ref{fig:heblue} that the spatial
distribution of the stars at young ages is not smooth, presenting a ``ring" shape. Also, the fast increase in the number of stars in the first three
panels in Fig. \ref{fig:heblue} (from left to right), suggests strong star formation activity in the past $\sim80$ Myr. 
However, the very last bin, which is close to 100 Myr, shows also
a particular spatial distribution, with a very low number of stars, and a different spatial distribution when compared with the previous panels. This difference
is observed on the CMD of the field 3 as a ``gap" between the RSG and the RGB (see Fig. \ref{fig:cmdsfh}). 
The velocity field presented by \citet{peters94} shows a clear disturbance at the location of the satellite
companion. We do not observe effects induced by the interaction (e.g. through tidal forces) on the CMDs/SFHs for the other two fields covered in this paper, however
our results are constrained up to $\sim$100 Myr due to incompleteness, which is a shorter time scale compared to the
time needed to complete the orbit suggested by Peters et al ($\sim$480 Myr). Our estimations showed a fast increase
in $\Sigma_{SFR}$ for the field 3 particularly (i.e. in the region not connected with NGC~1313). 
To fully understand what happened in the south-west region of NGC~1313, a numerical simulation will be needed.

In general, we do not see physical effects induced over the fields 1 and 2 by the passage of the satellite companion close to NGC~1313 (based on the estimated CMDs and SFHs), 
if we assumed that the companion galaxy followed the orbit suggested by \citet{peters94}. We can, in general, say
that the regions F1.a, F2.b, F1.c, and F2.a present the same levels of SFRs in the age range 10 to 100 Myr (see table \ref{tab:sfhs}), 
with no indication of disturbance by a tidal interaction (i.e. fairly constant). 
However, the regions F2.b, F3.a, and F3.b present similar levels in their SFRs, which are low compared to the ones observed in the other regions (see table \ref{tab:sfhs}).
The differences between the F3.a, F3.b and F2.b is that the former two present an increase in the last 100 Myr, while the later does not.

When comparing results expected from an interaction, such us gas dissipation, changes in shape and
enhance star formation \citep{schweizer00}, we agree with the scenario which states that NGC~1313 is
suffering an interaction with a satellite disrupted companion. This claim is based on the observed
increase in the $\Sigma_{SFR}$ in the south-east of the galaxy and the patchy shape observed in the northern arm of the galaxy.
Beside these observational evidence, the SFH in fields 1 and 2, do not show any changes, appearing
to be constant over the time span reached in this work. This result is interpreted as an interaction that did not
triggered a global star formation, but a local one.

\section{Summary and conclusions}
Using HST/ACS observations in the bands V and I we have studied the stellar population
of NGC~1313. Applying the synthetic CMD method \citep{tosi91} we estimated the star
formation histories over the past 100 Myr in three different regions of the galaxy.

We used the red and blue He burning phases as tracers of the star formation thanks to the
well age-dated signatures based on theoretical isochrones, and observed different
areas over the fields of NGC~1313. Field 1
shows a roughly constant activity in the arm, the bar and the H$\alpha$ regions (F1.a, F1.b, and F1.c). The second field
showed two separate group of stars, one of them (the region F2.b) showing to
be no connected with the bar or any of the arms of the galaxy. Field 3 showed 
a population that is not connected with the southern arm of the galaxy, but that has
experienced a recent burst of star formation.


The SFHs obtained from the study of resolved stellar populations in our field 3 showed an increase in the SFR, which can be interpreted as
an interaction process in the past of NGC~1313 with a tidally disrupted satellite companion. The SFH in the region
where the satellite companion is located shows an increase during the last 100 Myr, while the rest
of the galaxy (i.e. the bar and the arms) present a constant SFH during the same time range.

We therefore conclude that NGC 1313 is likely to have undergone an
 interaction with a minor companion that triggered a local starburst
 in the south-western part of the galaxy about 100 Myr ago, but was
 insufficient to strongly affect the global star formation history.
 Nevertheless, it is interesting to note that one of the most massive,
 young star clusters in NGC 1313 is also found in this region \citep{silvavilla11,larsen11}.
 
\section*{Acknowledgments}

We would like to thank the referee for comments that helped to improve this article.
This work was supported by an NWO VIDI grant to SL.

\end{document}